\documentclass[aps,prl,preprint,groupedaddress,showpacs,amsmath, amssymb, floatfix]{revtex4-1}
\usepackage{epstopdf}
\usepackage{graphicx}
\usepackage{color}
\usepackage{latexsym}
\usepackage{amssymb}
\usepackage{bm}
\usepackage{dcolumn}

\begin{document}

\title{Planar and Nematic Aerogels: DLCA and Superfluid $^3$He}

\author{M. D. Nguyen$^1$}
\email{mannguyen2019@u.northwestern.edu}
\author{J. S. Simon$^1$}
\author{J. W. Scott$^1$}
\author{Y. C. Cincia Tsai$^1$}
\author{A. M. Zimmerman$^2$}
\author{W. P. Halperin$^1$}
\email{w-halperin@northwestern.edu}
\affiliation{$^1$Department of Physics and Astronomy, Northwestern University}
\affiliation{$^2$Department of Physics, Harvard University}

\date{\today}

\begin{abstract}
We perform cluster aggregation simulations to model the structure of anisotropic aerogel. By biasing the diffusion process, we are able to obtain two distinct types of globally anisotropic aerogel structures which we call "nematic", with long strands along the anisotropy axis, and "planar", with long strands in planes perpendicular to the anisotropy axis. We calculate the auto-correlation function, the structure factor, and the angular dependence of the free-path distribution for these samples. 
The calculated structure factor from simulated aerogels can be compared with data from small-angle X-ray scattering (SAXS) of lab-grown aerogel allowing us to classify the spatial structure of the lab-grown samples. 
We find that the simulated "nematic" aerogel has a structure factor consistent with lab-grown, axially-compressed silica aerogel while the simulated "planar" aerogel has a structure factor consistent with lab-grown "stretched" silica aerogel.
Unexpectedly, compressing previously isotropic silica aerogel leads to the formation of long strands along the compression axis while stretching silica aerogel leads to formation of planes perpendicular to the stretching axis. We discuss the implication of this determination on experiments of superfluid $^3$He in anisotropic aerogel, in particular the orbital analog of the spin-flop transition. 
\end{abstract}

\maketitle

\section{Introduction}
Disorder and impurities are important in determining the relative phase stability \cite{imr.75} and engineering of novel phenomena \cite{fis.89} in a diverse set of superconductors and superfluids. In the high-T$_c$ cuprate supercondcutors, doping disorder strongly affects the prescence of superconductivity \cite{pan.01,lee.06,kei.15}, vortex physics \cite{ous.96}, and low-energy surface states \cite{yaz.99}. For superconducting radio frequency cavities used in particle accelerators, surface impurities improve the quality factor of the cavities \cite{gra.13,nga.19}. Recently, aerogels have been used to introduce correlated impurities in superfluid $^3$He to manipulate the phases \cite{hal.08,dmi.15,zhe.16} and stabilize new features such as half-quantum vortices \cite{aut.16}. Different types of aerogel structures and anisotropy will induce different properties in the superfluid \cite{vol.08,ask.15, li.15}. To better understand how aerogel structure affects these systems, we use diffusion-limited cluster aggregation (DLCA) simulations to model anisotropic aerogels.

There is extensive literature on using simulations to model globally homogeneous, isotropic aerogel (HIA) \cite{mea.83,kol.83,has.94,ma.01,det.03}, similar to those seen in Fig \ref{fig:hia}\textbf{a}. Here, we present a framework for generating, analyzing, and classifying aerogels with uniaxial anisotropy that have large scale structure not present in HIA. Finally, we use this classification to propose a mechanism for the recently observed orbital analog of the spin-flop texture transition of superfluid $^3$He \cite{zim.18}.

The DLCA simulations create an aerogel network by a process similar to that described by Hasmy \textit{et al.} in Ref. \cite{has.94} which we summarize in the following section. To obtain anisotropic aerogel, we modify this procedure by biasing the diffusion process along one axis defined to be the $\bm{z}$-axis. The degree of anisotropy is labeled by a single continuous variable, $\bm{\epsilon}=\epsilon\medspace\bm{\hat{z}}$, with $\epsilon$ defined to be the ratio of diffusivity along the $\bm{z}$-axis to the diffusivity perpendicular to $\bm{z}$. Isotropic aerogel has an anisotropy parameter of $\epsilon = 1$. Samples with $\epsilon > 1$ and $\epsilon < 1$ have markedly different large scale structures representing two classes of anisotropic aerogel. These structures have distinct signatures in their correlation functions and structure factors which can be directly calculated from the aerogel network. The structure factor is a particularly useful metric as it can be compared with small-angle x-ray scattering (SAXS) data obtained from aerogel materials \cite{has.94,nyg.12}. SAXS data for anisotropic aerogels show clear anisotropy in the scattering pattern but the underlying structure can not be determined because the scattering data is only proportional to the amplitude of the scattered wave with no phase information \cite{sto.89}. Therefore, it is not possible to fully reconstruct the underlying structure from the SAXS data alone \cite{leg.14}. On the other hand, starting from simulated aerogel with a well understood microscopic structure, we can calculate the structure factor and compare it with the SAXS data. We leverage this connection to classify real silica aerogel used in superfluid $^3$He experiments and demonstrate that the structures in the aerogel induce the orbital-flop transition \cite{zim.18}.

\section{Simulated Anisotropic Aerogel}
Aerogel can be accurately simulated using the procedure detailed in Ref. \cite{has.94}. A random point field of $N$ particles (ranging from N = 5000 to 200000) is initialized in a periodic box with volume $L^3$. The particles have a distribution of radii given by a log-normal distribution with a sample mean of $r_0$ and sample variance $\sigma_0$. All lengths in the simulation are normalized to $r_0$ yielding dimensionless distance parameters (such as $L/r_0$ which is used to characterize finite size effects). The variance was fixed at $\sigma_0/r_0 = 1/8$ because it does not affect the large scale structure for reasonable values of $\sigma_0$. The particles are allowed to diffuse randomly until they collide with another particle. If a collision occurs, the two particles are joined into an aggregate and thereafter diffuse together. The diffusion coefficient is controlled by the size (mass) of the aggregate, with larger clusters diffusing slower. When all particles are joined into a single cluster, the simulation ends yielding an aerogel cluster. 

\begin{figure*}
\includegraphics[width=1\textwidth]{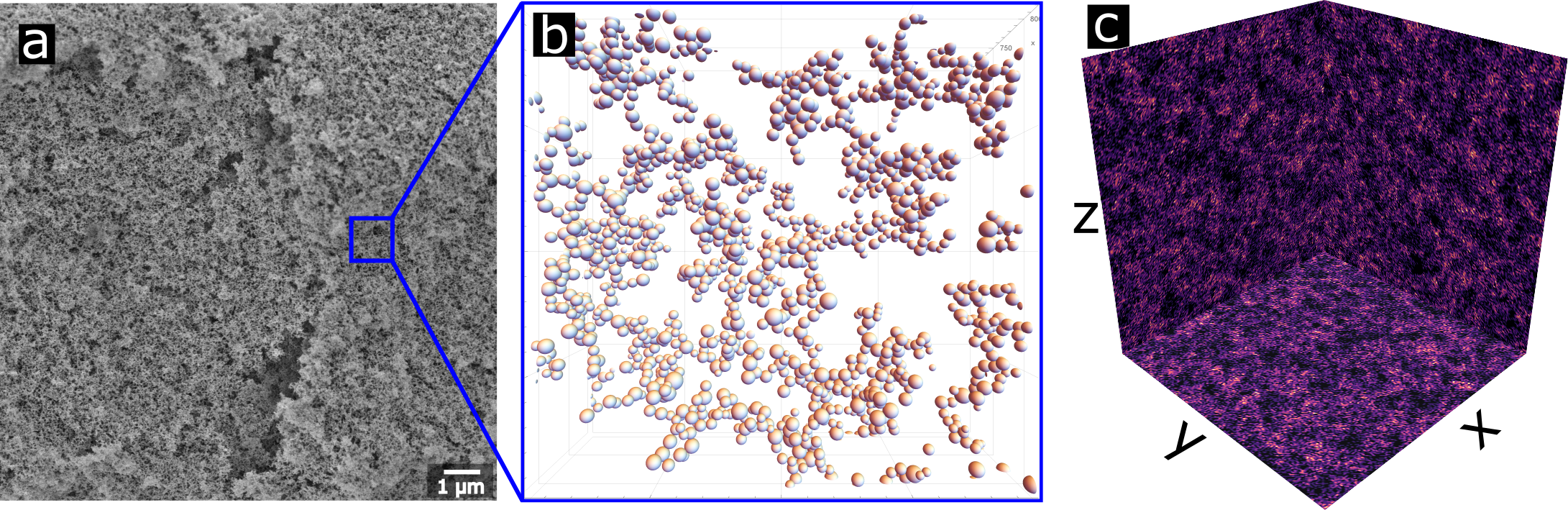}
\caption{Real and Simulated Isotropic Aerogel. \textbf{a} Scanning electron microscopy of real, $98\%$ porous isotropic aerogel shows the complex network of silica particles. \textbf{b} Simulated aeorgel cluster for isotropic diffusion for a small segment of the sample ($\sim 0.5\%$ of the total sample) (middle). Structural properties such as strand orientation, clustering, and void size are difficult to determine in the 3-D representation for the full sample. \textbf{c} Projecting the cluster onto orthogonal, 2-D planes reveals the position of silica spheres to be non-random. Each plane represents a projection of the aerogel sample along the axis perpendicular to that plane. For isotropic diffusion, the strands of silica appear to be without a preferred direction. However, characteristic cluster and void sizes are  visibly apparent.}
\label{fig:hia}
\end{figure*}

The resulting cluster is a density field denoted $\rho(\bm{r})$. For a discrete field of silica spheres, $\rho(\bm{r})$ is simply a list given by $\rho(\bm{r})=\lbrace \lbrace \varrho_1,\bm{r_1}\rbrace, ...,\lbrace \varrho_N,\bm{r_N}\rbrace \rbrace$, where $\varrho_i$ is the radius and $\bm{r_i}$ is center of the the $i^{th}$-particle. Fig. \ref{fig:hia} shows a sample density field for isotropic aerogel showing the complex aerogel structure and various particle sizes. Our work is focused on high porosity (low density) aerogel with the a filling fraction of $\rho_0\sim 2\%$, where $\rho_0=\frac{4}{3}\pi r_0^3\medspace\frac{N}{L^3}$, corresponding to real silica aerogel with mass density $\sim$ 45 mg/cm$^3$ \cite{zim.13}.

The full 3-D rendering of $\rho(\bm{r})$ obscures the strand and clustering of the aerogel network so 2-D, orthogonal projections are used to better visualize the spatial variation in the density field. The right-hand panel of Fig. \ref{fig:hia} shows the highly-correlated distribution of particle position, complex strand structure, and characteristic voids in the aerogel network. For the case of isotropic aerogel, all these properties have no preferred direction in space. In this work, we show that anisotropy can be introduced by biasing the diffusion step size along the $\bm{z}$-axis. $\epsilon >1$ indicates faster diffusion (larger step size) along the $\bm{z}$-axis while $\epsilon <1$ indicates faster diffusion in the $XY$-plane.

\begin{figure*}
\includegraphics[trim= 0 10 0 10, width=1\textwidth]{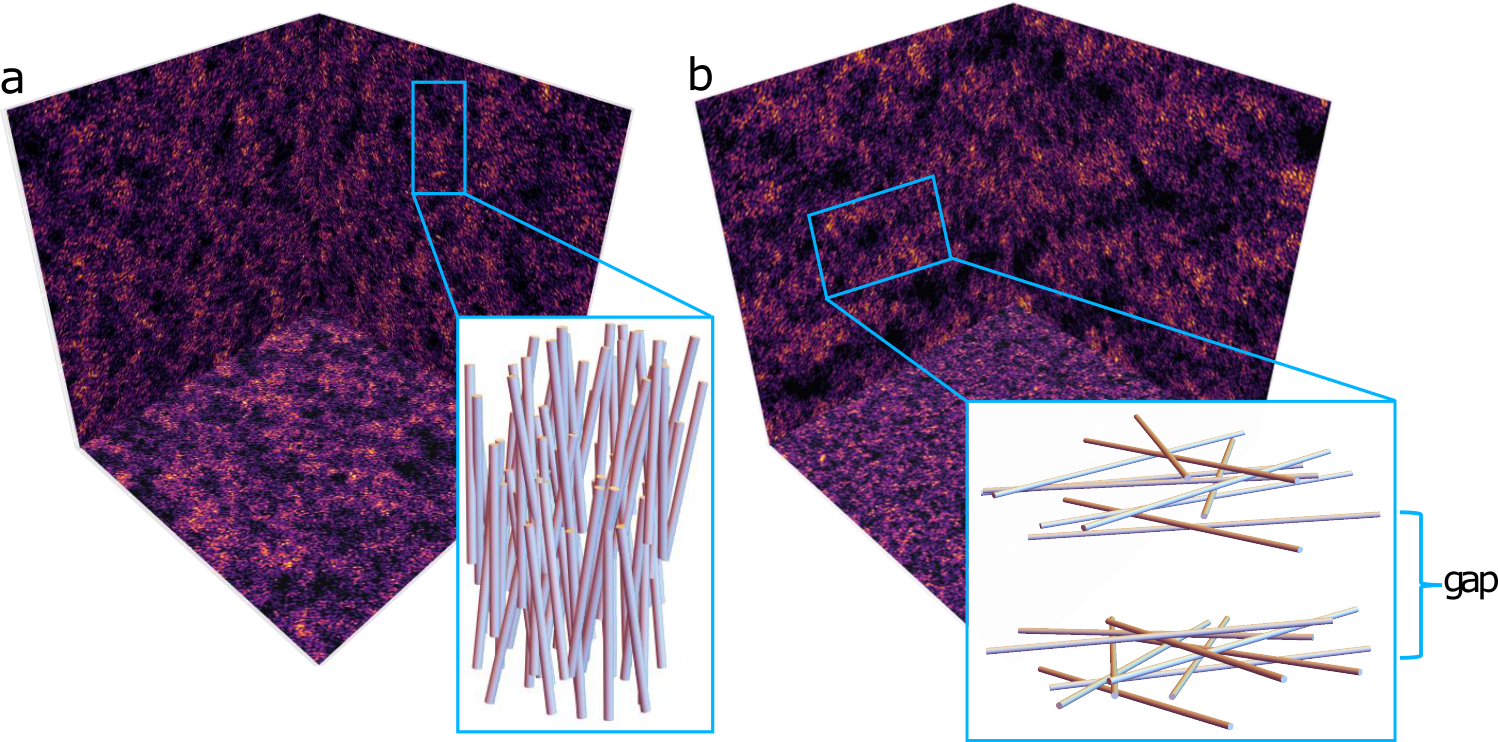}
\caption{Projection of aerogel structure for anisotropic aerogel.  The projections reveal clear anisotropic strand structure for $\epsilon \neq 1$ (as compared with isotropic aerogel in Fig. 1\textbf{c}). \textbf{a}  For $\epsilon = 0.25 $, the projections onto the XZ- and YZ-planes reveal long structures parallel to the anisotropy direction, $\bm{\epsilon}$, corresponding to nematic strands (inset). The projection along Z into the XY-plane reveals the strands oriented along Z are still correlated in their positions in the XY-plane. \textbf{b} For $\epsilon = 4 $, sheets of aerogel strands form in the XY-plane, perpendicular to $\bm{\epsilon}$ with gaps between sheets (inset). The projection along Z into the XY-plane reveals a random structure indicating that the orientation of the strands between distantly separated sheets are uncorrelated.}
\label{fig:cartoon}
\end{figure*}

For anisotropic aerogel, these projections reveal clear spatial anisotropy and large scale structure not found in the isotropic samples. We have numerically created two types of anisotropic aerogels with uni-axial, anisotropic diffusion which we classify as nematic, with $\epsilon < 1$, and planar, with $\epsilon > 1$. As seen in the the projected view of $\rho$ in Fig. \ref{fig:cartoon}, anisotropic diffusion introduces a preferred direction breaking the full 3-D rotational symmetry of isotropic aerogel. In the case of $\epsilon < 1$, the strands are preferentially aligned along the anisotropy direction $\bm{\epsilon}$. This is akin to nematic systems where long molecules have orientational order along one axis and absence of regular spatial ordering in the  perpendicular plane. For $\epsilon > 1$, the projected view along the $X$- and $Y$-axes reveals high-density, planar sheets of aerogel clustered together with some characteristic thickness. These sheets are separated from each other by visible gaps of low density regions with fewer particles. We classify samples with $\epsilon > 1$ as planar aerogels. A more quantitative description of these nematic and planar structures is obtained by calculating the autocorrelation function, the structure factor, and the distribution of geometric free paths. These three descriptors unambiguously differentiate between nematic and planar aerogels.


\section{Characterization}
\subsection{Correlation Function}
Silica spheres aggregate to form strands  which cluster together to form larger structures that make up the aerogel network. The positions of the spheres are non-uniform and highly correlated in space. This non-uniformity is encoded in the autocorrelation function, $g(\bm{r_i},...,\bm{r_j})$, which is the two-point characteristic of $\rho$ obtained by point-wise multiplication of $\rho$ evaluated at all pairs of points $\bm{r_i}$ and $\bm{r_j}$. For a globally homogeneous cluster with $N$ particles and volume $V$, $g$ only depends upon the separation, $\bm{R_{ij}}= \bm{r_i - r_j}$, given by
\begin{equation}\label{eq:xidef} 
g(\bm{R}) = \frac{
\langle\rho(\bm{r_i})\medspace\rho(\bm{r_j})\rangle 
}{N(N-1)/(2V)}
\end{equation}
where the  angled-brackets, $\langle...\rangle$, represent an ensemble average over all pairs (the homogeneous assumption) and the denominator $N(N-1)/(2V)$ is the mean density of pairs.
When normalized to the density of pairs, the autocorrelation function gives the excess likelihood to find two particles separated by a vector $\bm{R}$, relative to a random uniform Poisson point field of the same density \cite{has.94,set.21,bor.99}. The correlation function defined in Eq. (\ref{eq:xidef}) is also sometimes called the "pair correlation function", "pair distribution function", or "radial distribution function" depending upon the application \cite{bor.99,mcq.00,fre.86}. In the limit of large separation, $R\rightarrow\infty$, $g(\bm{R}) \rightarrow 1$, thereby indicating no excess correlation above that of a uniform distribution. 

While different samples drawn from the same probability distribution (i.e. simulated with the same parameters or experimentally grown under the same conditions) will have a different value for the density field $\rho(\bm{r})$ at any point $\bm{r_i}$, the two samples will have the same two-point functions because the correlations remain the same. Therefore, $g(\bm{R})$ can be averaged between samples while $\rho$ cannot. In addition, most applications of aerogel are not interested in the location of the silica spheres themselves but rather the open space between them, $i.e.$ the negative of the aerogel structure. Characteristic cluster and void sizes can be determined from the correlation function. Excess correlation ($g>1$) indicates clustering at that separation distance and direction while a deficit in correlation ($g<1$) indicates voids.

\begin{figure*}
\includegraphics[trim=0 0 0 100, width=1\textwidth]{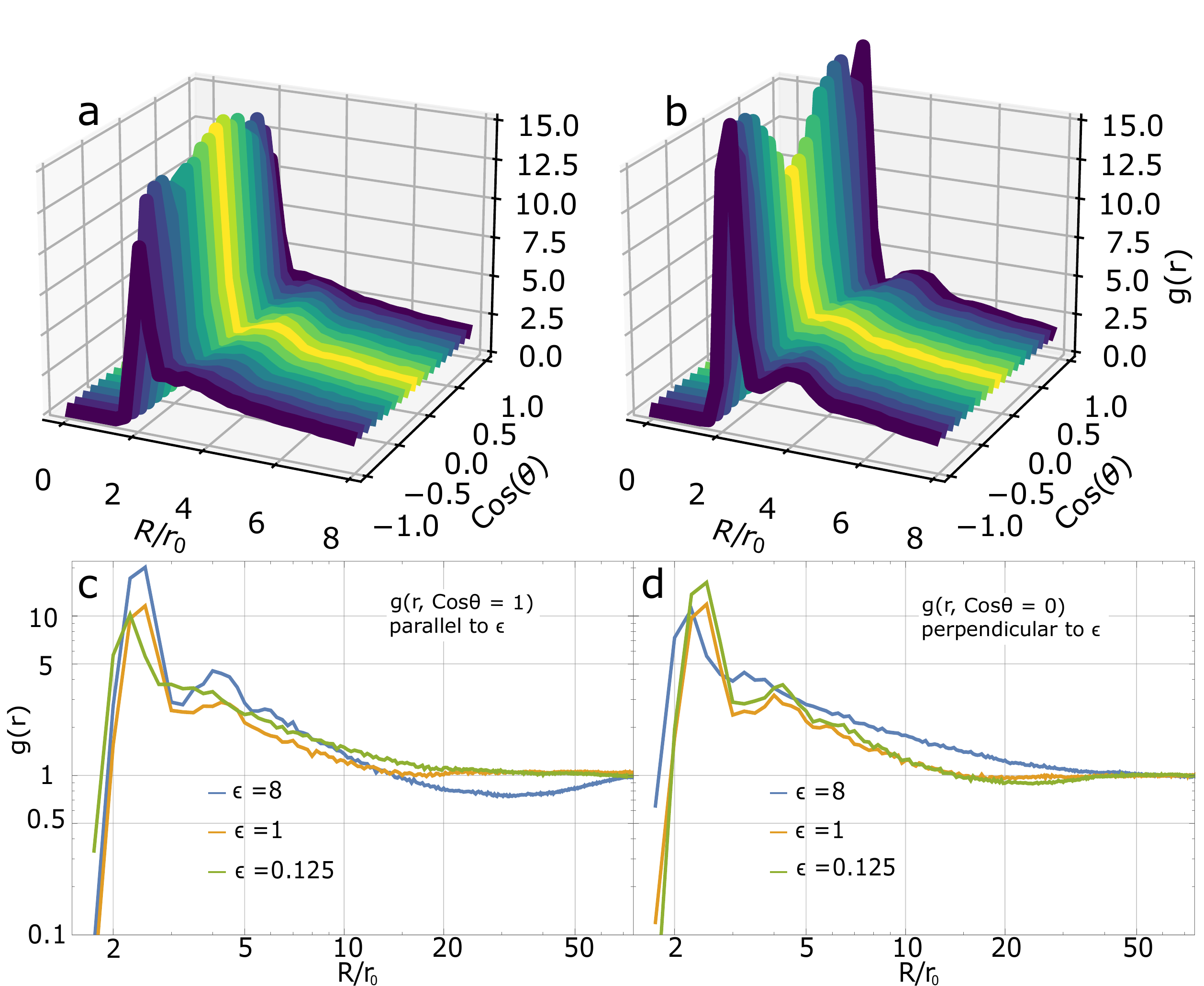}
\caption{Anistropic pair correlation of nematic and planar aerogels as a function of separation $R/r_0$ and cos$(\theta)$. \textbf{a} For nematic ($\epsilon = 0.25$) and  \textbf{b} for planar ($\epsilon=4$), where cos$(\theta) = \pm 1$ indicates pair correlations parallel to $\epsilon$ and cos$(\theta) = 0$ indicates correlations in the  perpendicular plane. The degree of excess and deficit in correlation can be tuned by changing $\epsilon$ as seen in the bottom two panels. \textbf{c} Pair correlations parallel to $\bm{\epsilon}$ for various values of $\epsilon$. Increasing $\epsilon$ increases the nearest-neighbor peak at short separations but also increases the deficit in correlation in the intermediate range of $10<R/r_0<50$. \textbf{d} Correlations perpendicular to $\epsilon$, for various $\epsilon$ values. Increasing $\epsilon$ in this direction has the opposite effect. The deficit in correlation in \textbf{d} is less than in \textbf{c}. This implies that the spacing between nematic strands for  $\epsilon<1$ samples is less than the gaps between sheets in the planar samples.}
\label{fig:cor_water}
\end{figure*}

For isotropic aerogels ($\epsilon =1$), $g(\bm{R})$ further simplifies to $g(R)$, depending only upon the magnitude of the separation. On the other hand, for anisotropic samples with $\epsilon \neq 1$, the angular-dependence of $g(\bm{R})$ becomes important. In the case of azimuthally symmetric, uniaxial anisotropy, $g(\bm{R})\rightarrow g(R,\theta)$, where $\theta$ is the polar angle with respect to the anisotropy axis $\bm{z}$. The correlation functions determined here have more structure than the correlation functions proposed in the literature which are simple power-laws with an upper fractal exponential cutoff \cite{fre.86}. This is not surprising as the aerogel is anisotropic with different macroscopic structure than isotropic aerogels. As seen in Fig. \ref{fig:cor_water}, $g(R,\theta)$ for nematic and planar aerogel have non-uniform $\theta$-dependence, a clear signature of anisotropy. The $R$-dependence of the deficit in correlation gives the characteristic size of voids and while the $\theta$-dependence of the deficit gives the shape of voids for each sample. In all direction, there is a significant nearest-neighbor peak around 2\,$r_0$ indicating contact between particles. The relative height of the nearest-neighbor peaks in different directions indicate whether pairing along $\bm{\epsilon}$ (cos$(\theta) = \pm 1$) or in the plane perpendicular (cos$(\theta) = 0$) is more likely. For nematic samples (the green curves in panel \textbf{c} and \textbf{d}), there is greater likelihood for the nearest neighbor to be in the plane perpendicular to $\bm{\epsilon}$ than parallel to $\bm{\epsilon}$ as seen in  Fig. \ref{fig:cor_water}\,\textbf{a}. However, at intermediate separation, 10\,$r_0 < R < 50 \,r_0$, the direction of excess correlation swaps, indicating particles are more likely to be collimated along $\bm{\epsilon}$. This is the signature of the long nematic strands. 

The opposite behavior is observed for planar ($\epsilon >1$) samples (the blue curves in panel \textbf{c} and \textbf{d}). At small separations, there is preferential pairing along $\epsilon$ but for larger separations, a neighbor is more likely to be found in the plane. Increasing $\epsilon$ increases the nearest-neighbor peak at short separations but also increases the deficit in correlation in the intermediate range of $\sim 20\,r_0$. This is interpreted to be the scale of the thickness of the planes of aerogel strands. A silica sphere located in the planes is less likely to observe a neighbor above or below it at distances greater than the sheet thickness but less than two sheet thickness.  Correspondingly, as $\epsilon$ increases, so does the size of gaps between the sheets for planar aerogel. In both the $\epsilon >1$ and $\epsilon <1$ cases, it is the behavior of the correlation function at the intermediate length scale of 10 to 50\,$r_0$ that is central to understanding the macroscopic properties of the aerogel. Visually, the correlation function is dominated by the nearest-neighbor peak at 2\,$r_0$, but this only describes the smallest scale correlation. 


There are two methods of numerically calculating $g(\bm{R})$, the "direct" method and the Fourier-Correlation method, each with their own advantages and disadvantages. The direct method simply applies the definition of the correlation function and loops through every pair of particles, calculates their separation vector $\bm{R}$, and histograms the set of $\bm{R}$ into equal volume $R$ and $\theta$ bins. The raw bin counts are then normalized by the density of pairs and the bin volume, $\frac{N(N-1)}{2V} 2\pi r^2 dR\,d($cos$\theta)$ for radial bin width $dR$ and theta bin width $d($cos$\theta)$. This method can be implemented natively in spherical coordinates but is slow in time, scaling as $\mathcal{O}(N^2)$ for $N$ particles. 

Because of finite sample size, particles near the edge of the sample box will have an artificial deficit in neighbors leading to the tail of the distribution (large $\bm{R}$) being incorrect. This can be corrected by several different methods as described in Ref. \cite{bor.99}. Astronomers calculating auto-correlation functions of galaxies observe biases for large $\bm{R}$ \cite{dav.83} and have devised various estimators to correct for this. The central idea is to consider a randomly distributed sample of similar size and density. This artificial sample will have the same finite size effects as the simulation of interest. The autocorrelations of the simulation of interest and of the artificial sample along with the cross-correlation \textit{between} the artificial sample and the simulation are combined to remove finite size effects. Different combinations of autocorrelations and cross-correlation have been suggested each with their own statistical bias \cite{bor.99}. We have implemented several of the most widely used estimators and compared them with a simple procedure of cross-correlating the original aerogel sample $\rho(\bm{r})$ with a copy of itself that has been spatially-shifted in all three directions by the simulation box size, $\rho(\bm{r}\pm L(\hat{\bm{x}},\hat{\bm{y}},\hat{\bm{z}}))$. This procedure is equivalent to applying the periodic boundary used in the simulation. We find that the latter method effectively corrects the tail of $g(\bm{R})$ consistent with the other estimators without the need to generate a test sample. The correlations in Fig. \ref{fig:cor_water} were determined using the direct method with periodic boundary conditions.

The second method uses the Fourier-Correlation (Wiener-Khinchin) theorem and the efficiency of fast-Fourier transforms (FFT) to speed up the process. The autocorrelation function can be obtained by first converting $\rho(\bm{r})$ into a sparse 3-D matrix, $\rho_{ijk}$, whose indices form a lattice and whose matrix elements represent the density at lattice site $(i,j,k)$. This matrix $\rho_{ijk}$ is numerically Fourier transformed into its conjugate field $\hat{f}_{lmn}$. Then the cartesian, pair-distribution function, $g_{xyz}$, is the inverse Fourier transform given by
\begin{equation}\label{fft}
g_{xyz}=\frac{\langle\rho_{ijk}\medspace\rho_{i'j'k'}\rangle}{N(N-1)/(2V)}=\frac{\mathcal{F}^{-1}\{ |\hat{f}_{lmn}|^2\}}{N(N-1)/(2V)}
\end{equation}
This calculates the circular autocorrelation of $\rho(\bm{r})$ which naturally enforces the periodic boundary used in the simulation so it does not have to be corrected for finite size effects. Due to the speed and efficiency of FFT algorithms, this method is significantly faster in time and scales only as $\mathcal{O}(N Log N)$. However, this method is memory intensive as the matrix representation of $\rho$ grows as $L^3$ for $L$ lattice sites in each dimension. For a cubic lattice with $10^3$ sites per dimension, $\rho_{ijk}$, stored as a 32-bit float will be $\sim$4 GB of data. In computational complexity, the FFT-correlation method scales well in time but poorly in space (memory), while the direct method is the opposite, efficient in space but slow in time. 
Importantly, the FFT method also calculates the three-dimensional, cartesian structure factor, $S_{xyz}$.

\subsection{Structure Factor}

From the structure factor, a connection can be made between simulated and real aerogel. The x-ray scattering intensity, $I(\bm{q})$, can be decomposed as $I(\bm{q})\propto S(\bm{q}) F(\bm{q})$. $F(\bm{q})$ is the single-particle form-factor that encodes details about the particle shape which affects the large-$q$ behavior of $I(\bm{q})$. $S(\bm{q})$ encodes correlation in position of particles at intermediate and large spatial scale (small-$q$). For small-angle x-ray scattering (SAXS), $I(\bm{q})$ is dominated by the $S(\bm{q})$ contribution. Therefore, the structure factor can be used to directly compare with SAXS data.

Again, there is extensive literature on determining $S(\bm{q})$ by calculating $g(\bm{R})$ and performing a Fourier transform \cite{has.94,fre.86}. This is usually done in spherical coordinates where the integration kernel simplifies from $e^{i \bm{q}\cdot\bm{R}}$ to sin$(qr)/(qr)$ because the aerogel under consideration is isotropic. 
In the case of an anisotropic $\rho(\bm{r})$, it is easier to numerically perform this Fourier transform in cartesian coordinates, then convert back to spherical coordinates. In fact, the structure factor is actually calculated first as an intermediate step when employing the Fourier-correlation theorem to determine $g_{xyz}$. The cartesian $S_{x y z}$ is given by \cite{siv.11}
\begin{equation}
S_{x y z} = |\mathcal{F}\{\rho_{xyz}\}|^2=|\hat{f}_{lmn}|^2.
\end{equation}
To compare with SAXS data for x-rays incident perpendicular to $\bm{\epsilon}$, $S_{x y z}$ is then converted to $S(\bm{q}_\parallel,\bm{q}_\perp)$, where $\bm{q}_{\parallel}$ is the component parallel to $\bm{\epsilon}$ and $\bm{q}_{\perp}$ is the component perpedicular.

\begin{figure}
    \centering
    \includegraphics[width=1\linewidth]{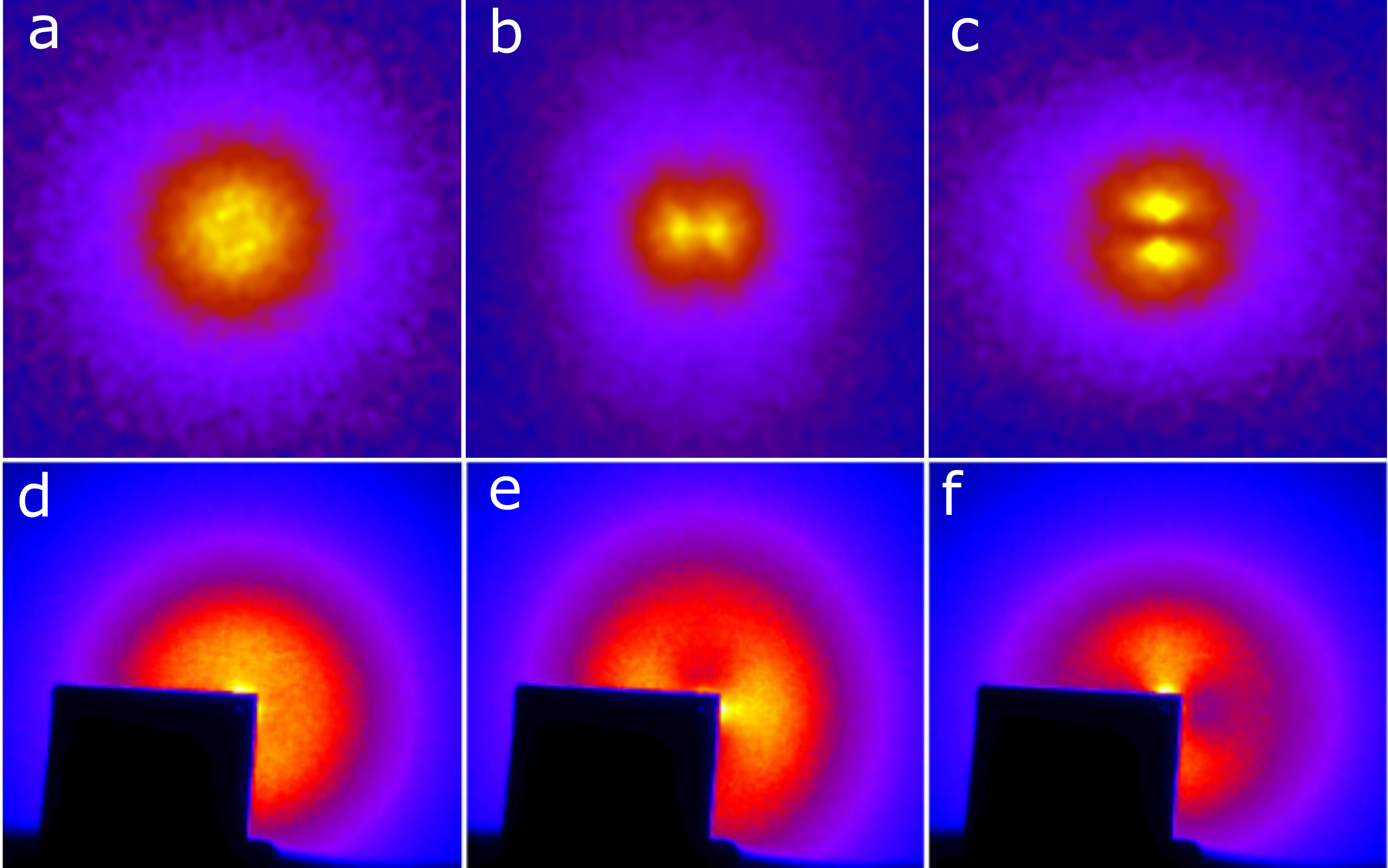}\\
    \caption{Calculated $S(\bm{q})$ versus Small-Angle X-ray Scattering Data \cite{pol.08a}, with the anisotropy axis vertical. The top panels display the calculated structure factor $S(\bm{q})$. \textbf{a} isotropic; \textbf{b} nematic ($\epsilon = 0.25$), and \textbf{c} planar ($\epsilon = 4$). For the nematic simulation, at short $q$, $S(\bm{q})$) has a dipolar shape with the "long axis" perpendicular to $\bm{\epsilon}$. For the planar case, at short ${q}$, $S(\bm{q})$ has a dipolar shape with the "long axis" along the anisotropy direction. 
    However, at larger {$q$}, as seen in the purple regions, the "long axis" of the anisotropy pattern is rotated by $90^o$.
The bottom panels display the Small-angle X-ray scattering (SAXS) for lab-grown \textbf{d} isotropic, \textbf{e} axially compressed (12.7 $\%$ negative strain), and \textbf{f} stretched (13.7 $\%$ positive strain) aerogel samples from Ref. \cite{pol.08a}. The black square in the data images is the beam stop.}
    \label{fig:struct}
\end{figure}

The calculated structure factor of simulated aerogel in Fig. \ref{fig:struct}\,\textbf{a,\,b,\,c} are compared with the SAXS data from real aerogel Fig. \ref{fig:struct}\,\textbf{d,\,e,\,f}. Isotropic aerogel with $\epsilon =1$ has the expected isotropic $S(\bm{q})$. For anisotropic samples, the structure factor has two features of note. First is the distinct dipolar angular distribution pattern at short ${q}$ (corresponding to a length scale of $\sim$100\,$r_0$) as seen in yellow and orange in panels $\textbf{b}$ and $\textbf{c}$. Secondly is the ellipsoidal pattern at intermediate ${q}$ (corresponding to $\sim$10\,$r_0$) as seen in the purple regions. The orientation of these two patterns can be used to unambiguously classify aerogels. Nematic-like aerogels will have the short-${q}$, dipolar pattern perpendicular to the anisotropy while planar-like aerogels will have the dipolar pattern parallel to $\bm{\epsilon}$. This is a general framework for classifying aerogel and will hold true irrespective of the material or type of aerogel. The orientation of the anisotropy of the structure factor is a general feature encoding the difference between nematic and planar correlations. 
At larger $q$ (in the purple region of panels \textbf{b} and \textbf{c}), the major ("long") axis of the ellipsoidal pattern is rotated $90^o$ from the short-$q$ dipole pattern. Evidently, there are two scales of structure for the aerogel. The large scale structure is reflected in the dipole scattering pattern and the smaller scale structure oriented perpendicular to the large structure is reflected in the ellipsoidal pattern of the SAXS and structure factor. We use the large scale behavior to classify and label the aerogel samples as being either nematic or planar. 


 Fig. \ref{fig:struct}\,\textbf{d,\,e,\,f} show the SAXS data for real aerogels. The two types of anisotropic aerogel analyzed are obtained by either compressing (negative strain) or stretching (positive strain) isotropic aerogels \cite{pol.08a,zim.13}. It was not previously known how this strain affected the underlying structure of aerogel. Comparing the SAXS data to our calculated $S(\bm{q})$, we can determine if compressing aerogel creates nematic or planar structure. Compressed aerogels, seen in panel \textbf{e}, has the dipolar scattering pattern at short ${q}$ with the "long" dipole axis perpendicular to the anisotropy axis. For stretched aerogels in panel \textbf{f}, the short ${q}$ dipole pattern is parallel to $\bm{\epsilon}$ while for intermediate ${q}$, the "long" axis is perpendicular to $\bm{\epsilon}$. In other words, the direction in which scattering is more intense rotates by $90^o$ as ${q}$ increases (going to smaller length scale), which is the same behavior observed in the calculated structure factors.

Comparing these SAXS patterns to the structure factor of simulated aerogels, we can identify the structure of experimentally produced aerogels. Axially compressed aerogel has a scattering intensity consistent with nematic aerogel. On the other hand, experimentally stretched aerogel is consistent with planar aerogel. Compressing isotropic aerogel unexpectedly leads to the formation of strands \textit{along} the compression axis. This is contrary to speculations from Volovik \cite{vol.08}. In his model, stretching aerogel is argued to create long nematic strands parallel to the stretching while compressing aerogel would collapse the strands into planes. Here we find the opposite behavior. Our identification however is consistent with experimental results of superfluid $^3$He imbibed in anisotropic aerogels\cite{pol.12b,li.14,zim.18,dmi.20} which we detail in the final section.


\subsection{Free Path Distribution}
\begin{figure*}
\includegraphics[width=1\textwidth]{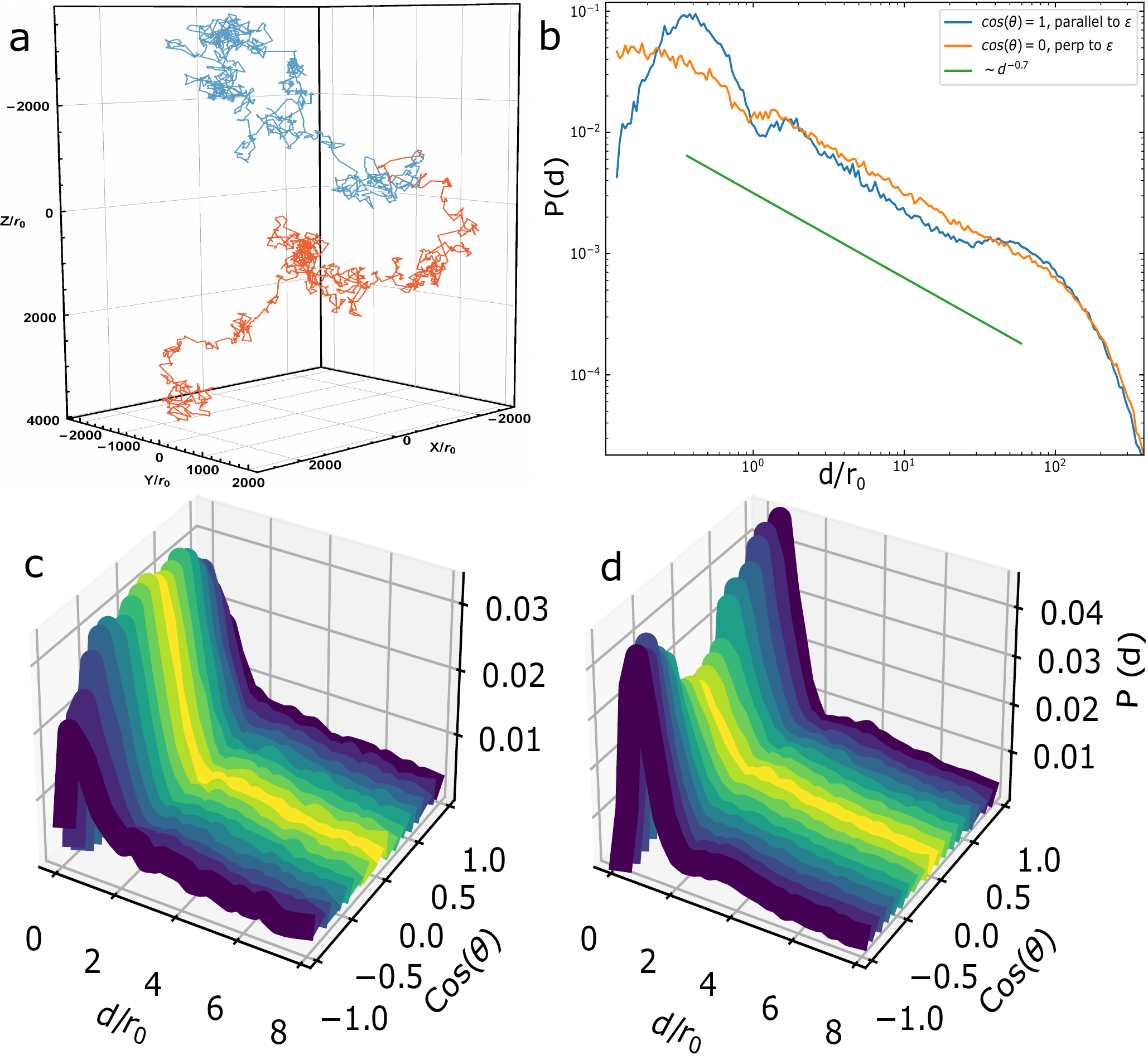}
\caption{Geometric free paths in the aerogel network. \textbf{a} Two free flights through nematic ($\epsilon = 0.25)$ aerogel starting at the same location at different angles. \textbf{b} Distribution of free path for planar ($\epsilon=8)$ aerogel parallel (blue curve) and perpendicular (orange curve) to $\bm{\epsilon}$. There is a power-law path distribution ($\sim d^{-0.7}$) for short paths before being exponentially cut off above 100\,$r_0$. 
The bottom panels show the full $P(d,\theta)$ for ($\epsilon=0.25)$  (panel \textbf{c}) and ($\epsilon=4)$  (panel \textbf{d}) with a similar nearest-neighbor angular dependence to the correlation function, Fig. \ref{fig:cor_water}}
\label{fig:mfp}
\end{figure*}

The previous two metrics, the correlation function and structure factor, characterize the aerogel structure itself. For many applications of aerogel, it is the void between the silica particles that is relevant rather than the aerogel network. An important measure of this negative space is the distribution of geometric free path through the aerogel (which appears as a parameter in  theoretical calculations of properties of $^3$He in aerogel \cite{zen.95,thu.98}). That is to say, starting at the surface of a random particle, how far can a test ray move before colliding with the aerogel network? The condition for collision between a ray and a sphere is given by the discriminant \cite{hai.89}:
\begin{equation}
disc=(\bm{\hat{d}}\cdot(\bm{p}_f-\bm{p}_i))^2 - (|\bm{p}_f-\bm{p}_i|^2-r_f^2) \geq 0
\end{equation}
where $\bm{d}=d\medspace\bm{\hat{d}}$ is the ray, $\bm{p}_i$ is the origin of the ray, and $\bm{p}_f$ is the center of the final sphere with a radius $r_f$.
If $disc\geq0$ and $\bm{\hat{d}}\cdot(\bm{p}_f-\bm{p}_i)\geq 0$ (this second condition ensures only collisions in the forward direction are considered), then the path length, $d$, is calculated as $d=(\bm{\hat{d}}\cdot(\bm{p}_f-\bm{p}_i))-\sqrt{disc}$. The free path is determined by taking the minimum $d$ observed along the direction of travel. If no collision is observed within the initial box, periodic boundary conditions are applied. The bounding box plane that the ray intersects is determined, and then the aerogel sample is shifted in the appropriate direction and collision detection is applied for the shifted sample. This is repeated until a collision is found. A probability density function, $P(\bm{d})$, is obtained by taking a histogram of the catalog of free paths. A random walk through the aerogel will have a distribution of step sizes given by $P(\bm{d})$.  Fig. \ref{fig:mfp}\,\textbf{a} shows the diverging path of two random walks through nematic ($\epsilon=0.25$) aerogel starting at the same particle but at different angles. These random walks exhibit the key feature of what are called "L\'evy flights" \cite{bar.08,vis.08}. The characteristic "jumps" of a L\'evy flight are observed where the test ray is confined to small regions followed by big jumps to other regions \cite{vis.08}. 

From $P(\bm{d})$, we can calculate a mean free path, $\lambda$, which has been shown to be inversely proportional to density, for low density samples \cite{haa.00}. We find that while the distribution of free path is very different for high porosity aerogel compared with a uniform Poisson point field of the same density, the mean free path for both systems are similar to within $7\%$ of 60\,$r_0$. There are two reasons for this. First, at low density, both a highly correlated system like aerogel and an uncorrelated uniform distribution will have large lines of sight. The excess correlation of the aerogel structure, which affects the distribution at short path lengths, has a smaller effect than density variations. Secondly, aerogel ceases to be a fractal above what is call the "upper-fractal cutoff" \cite{fre.86}. If aerogel had no upper fractal cutoff, then the free path distribution will be scale-free and described by a generalized L\'evy distribution (the namesake of the L\'evy flight) with the asymptotic form $P(d) \sim d^{-\alpha}$, with $1<\alpha<3$ \cite{bar.08,vis.08}. This power-law distribution is fat-tailed meaning there is significant weight of the distribution in long free paths with the possibility that the mean is undefined (for $\alpha\leq2$). 

However, both SAXS data and theoretical calculations show that aerogel is not fractal at all lengths \cite{fre.86,pol.08a}. Correspondingly the free path distribution is cut off and the mean is well defined. These "truncated L\'evy flights" however still retain many properties of L\'evy flights such as super-diffusion \cite{man.94}. As seen in Fig. \ref{fig:mfp}, the distribution is indeed power-law below 100 $r_0$ with a very weak exponent of $\alpha =0.7$ indicating a very flat probability distribution. At longer length scales above $\sim 100\,r_0$, the distribution is exponentially cut off. The cutoff is not due to finite size effects of the simulation as it remains constant with increasing $L/r_0$ from 100 to 350. 

For uniaxially anisotropic aerogels, $P(\bm{d}) \Rightarrow P(d,\theta)$, being a function of both path length, $d$, and polar angle, $\theta$. The height of the peak of the distribution has a $\theta$-dependence similar to what is observed for the correlation function in Fig. \ref{fig:cor_water}. Also like the correlation function, the behavior of $P(d,\theta)$ for intermediate distances $\sim$\,20\,$r_0$ is different from short distances. For planar aerogel, Fig. \ref{fig:mfp}\,\textbf{b}, there are more free paths along $\bm{\epsilon}$ at very short distances but more free paths perpendicular to $\bm{\epsilon}$ at intermediate distances, consistent with the existence of large planar gaps in the structure as indicated in Fig. \ref{fig:cartoon}\,\textbf{b}.

Each angle can be considered an independent probability distribution and distribution moments can be defined at different angles. Two directions of particular interest are the mean free path parallel, $\lambda_\|$ , and mean free path perpendicular, $\lambda_\bot$, to the anisotropy direction $\bm{\epsilon}$. Despite very clear anisotropy, the first moments of $P(d,\theta)$ (mean free path along a certain direction) are similar for each of the two orthogonal directions. For $\rho_0\sim 2\%$, $\lambda_\| \sim \lambda_\bot \sim 60\,r_0$ in both the nematic and planar aerogels. For real silica aerogels used in superfluid $^3$He experiments, $r_0$ is $\sim 1.5-2$ nm, indicating a mean free path of $\sim 90-120$ nm \cite{thu.98}, and can be much larger, up to $\sim$10\,nm in general \cite{fri.92,cai.20}), consistent with experimental measurements for isotropic aerogel of comparable density \cite{bak.13}. Consequently, the mean is not a good parameter to characterize the path distribution of high porosity anisotropic aerogel. In addition, the two length scales in the anisotropic samples are further hidden when only the mean free path is considered. The blue and orange curves in Fig. \ref{fig:mfp}\,\textbf{b} are quite different and yet they have the same first moment. We conclude that it is insufficient to simply use the mean free path to encode the effects of anisotropic aerogel in theoretical calculations.

To recap, we find that the correlation function, structure factor, and distribution of free paths form a set of metrics that can be used to characterize and classify anisotropic aerogels. From this process, it was determined that axially compressed silica aerogel has nematic strands while stretched aerogel has planar structure. We use this determination in the following section to explain a set of superfluid $^3$He experiments that employ these aerogels. 

\section{Anisotropic Aerogel and Superfluid $^3$He}
Anisotropic aerogels have recently found use in superfluid $^3$He where it stabilizes novel order parameter structures such as half-quantum vortices, superfluid polar-phase, and the orbital angular momentum analog of the spin-flop transition in antiferromagnets \cite{aut.16,zhe.16,li.13,zim.18}. Here we propose a mechanism for the "orbital-flop" transition based on our identification of different length scales present in anisotropic silica aerogel. 

\begin{figure}
\includegraphics[width=0.7\textwidth]{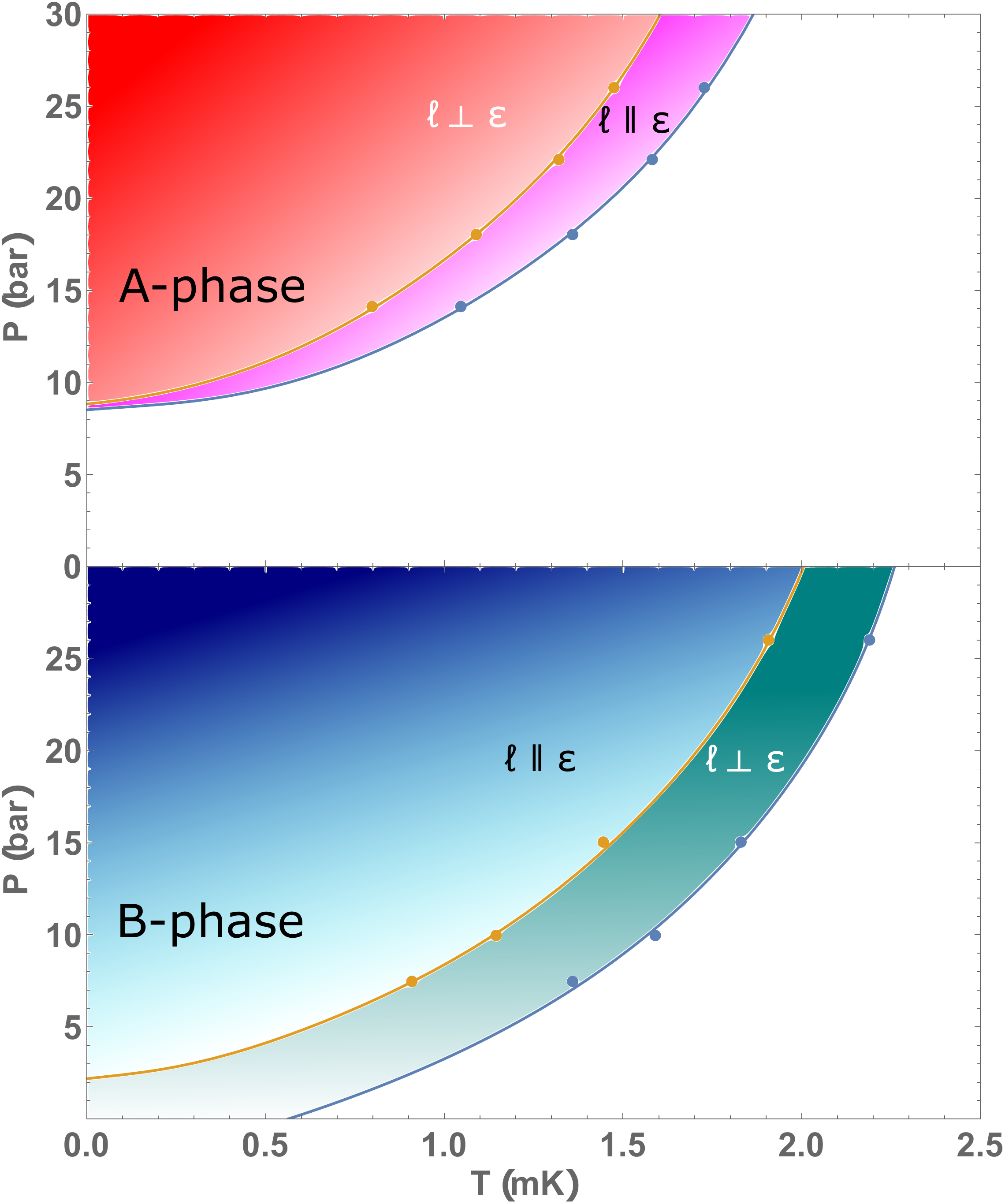}
\caption{Phase diagram of superfluid $^3$He in 14 $\%$ stretched aerogel (top) and 20 $\%$ compressed aerogel (bottom) extrapolated to zero magnetic field. In stretched aerogel, there is only the A-phase while there is only the B-phase in compressed aerogel. The orientation of $\ell$ is reversed in the two systems.}
\label{fig:phasediag}
\end{figure}

Superfluid $^3$He is an unconventional, topological superfluid with quasiparticles forming $p$-wave (L = 1), spin-triplet (S = 1) Cooper pairs creating a manifold of possible phases. In the chiral "A-phase", the Cooper pairs have a net orbital angular momentum, $\bm{\ell}_A$, with a vector order parameter. In the isotropic "B-phase", the Cooper pairs exist in a superposition of all three components of spin and orbital angular momentum projections with total angular momentum $J = 0$. The relative stability of the phases is strongly affected by aerogel. In the pure superfluid, both the A and B-phases can exist as stable equilibrium phases depending upon temperature, pressure, and magnetic field. This phase diagram is drastically altered in the presence of anisotropic aerogel. For compressed aerogel in zero magnetic field, only the B-phase is observed for the entire pressure and temperature phase diagram \cite{zim.18,zim.20}. On the other hand, for stretched aerogel, the A-phase becomes the equilibrium phase at all magnetic fields, pressure and temperature \cite{pol.12b,li.13}. In addition to altering the stability of phases, anisotropic aerogel has been observed to reorient the orbital degrees of freedom \cite{zim.18}. Fig. \ref{fig:phasediag} shows phase diagrams of these two systems.

In the presence of symmetry breaking effects such as magnetic fields, boundaries, or anisotropic disorder, the B-phase becomes distorted in its orbital degrees of freedom giving rise to a preferred direction denoted $\bm{\ell}_B$. Recently, sharp transitions have been observed where the orbital vectors in the two phases spontaneously reorient by $90^\circ$ uniformly across the entire system as temperature or pressure is varied \cite{li.14,zim.18}. It was determined that this reorientation is dependent upon the anisotropy of the aerogel and not from competing orienting effects such as from boundaries as has been
observed in isotropic aerogel [49].

\begin{figure}
\includegraphics[width=1\textwidth]{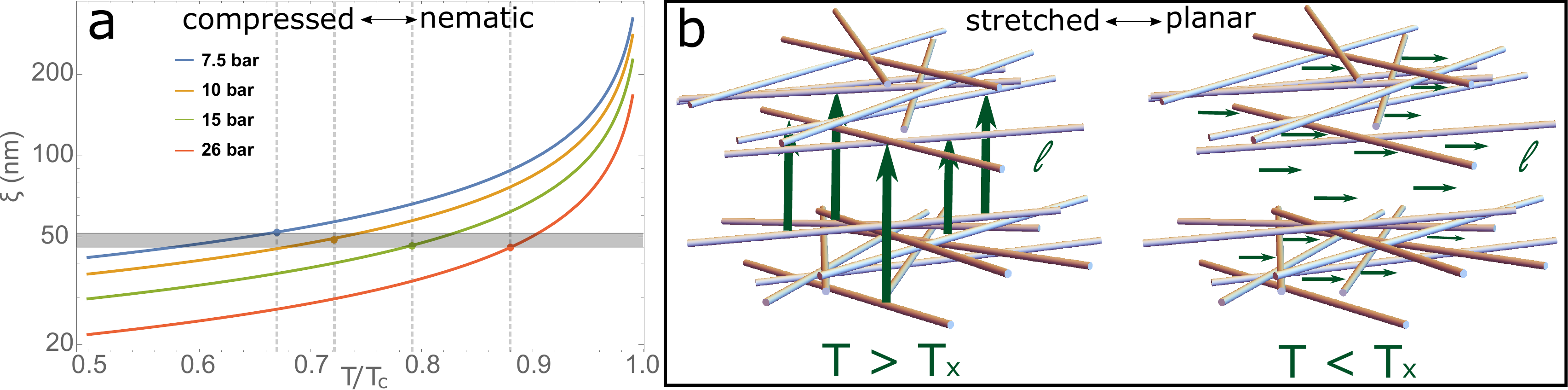}
\caption{\textbf{a}, The Ginzburg-Landau coherence length $\xi(T,P)$ for various pressures. The orbital flop transition $T_x$ in the B-phase of compressed aerogel for each pressure is indicated by the data point and the vertical dashed line. $T_x/T_c$ varies with pressure but the coherence length evaluated at  various $T_x$, $\xi(T_x,P)$, all fall into a narrow band around 49 nm. This indicates that the orbital flop transition  occurs when the superfluid coherence length decreases that length scale. \textbf{b}, Orientation of the orbital angular momentum in stretched aerogel which has been identified as planar aerogel. Above $T_x$, the coherence length is large and $\bm{\ell}$ is oriented perpendicular to the large scale planar structure. Below $T_x$, the coherence length is small and $\bm{\ell}$ reorients to being perpendicular to the small scale structure.}
\label{fig:xi}
\end{figure}

Phase identification of the superfluid,  and identification of the direction of the angular momentum axis can be determined from NMR spectra obtained in a high homogeneity steady magnetic field,  discussed most recently by Zimmerman $et\,al.$ \cite{zim.19}  In the superfluid A-phase of stretched aerogel, $\bm{\ell}_A$ orients parallel to the anisotropy axis $\bm{\epsilon}$ at high temperature near the superfluid transition, $T_c$. NMR experiments \cite{pol.12b,li.14} show that at a lower temperature denoted $T_x$, $\bm{\ell}_A$ spontaneously flops over to being perpendicular to $\bm{\epsilon}$ across the entire sample, as depicted in panel  of Fig. \ref{fig:xi}\,\textbf{b}. In the superfluid B-phase of compressed aerogel, $\bm{\ell}_B$ is initially perpendicular to $\bm{\epsilon}$ near $T_c$, the opposite of what is observed in the A-phase of stretched aerogel. At $T_x$, $\bm{\ell}_B$ sharply reorients to being parallel to $\bm{\epsilon}$ with a narrow transition width of $\sim$ 15\,$\mu K$ \cite{zim.18}. This orbital-flop transition varies with pressure between $\sim$ 0.67 $T_c$ at 7.5 bar to 0.88 $T_c$ at 26 bar. The opposite behavior of these two samples is resolved by considering the underlying structure of the aerogel. 


From both the SAXS data and $S(\bm{q})$, it is evident that these anisotropic aerogels have different structure at long and short length scales. The large scale structure is given by the dipolar pattern while the small scale structure is given by the ellipsoidal pattern at large $\bm{q}$, Fig. \ref{fig:struct}. Furthermore, the scattering patterns reveal that large scale structure is oriented perpendicular to the small scale structure. 
We propose that this structural crossover in the aerogel induces the orbital-flop transition. The most important length scale in a superfluid is the coherence length, $\xi$, which can be thought of as the size of a Cooper pair (or more accurately, the healing length for variations of the order parameter). The coherence length is largest near the superfluid transition and decreases with temperature. Therefore, at high temperature near $T_c$, the superfluid's orbital degrees of freedom will be sensitive to large scale disorder. As $\xi$ becomes smaller at lower temperature, the smaller scale structure in the aerogel dominates.

The analysis presented in this work unambiguously identifies that at long length scales stretched aerogel has planar structure while compressed aerogel has nematic structure . For planar aerogels, the surface normal of the large scale structure points along the anisotropy axis,  $\bm{\hat{\epsilon}}$. Correspondingly in the A-phase of superfluid $^3$He in planar aerogel we would expect $\bm{\ell}_A\| \bm{\hat{\epsilon}}$ at high temperatures above $T_x$, and $\bm{\ell_a}\bm{\perp} \bm{\hat{\epsilon}}$ below $T_x$ \cite{pol.12b,li.14}. If the preferred orientation of $\bm{\ell}$ is determined solely by aerogel structure, it must be independent of the superfluid phase.  Consequently for a B-phase in nematic aerogel, parallel and perpendicular orbital orientations are just interchanged as seen in Ref. \cite{li.14,zim.18}. Above and below $T_x$, $\bm{\ell}$ preferentially orient perpendicular to the dominant aerogel structure. If this is the mechanism for the transition, we expect that the coherence length evaluated at $T_x$, $\xi(T_x,P)$ to be relatively constant at different pressures. The transition occurs at $T_x$ because that is temperature at which the superfluid becomes more sensitive to the small scale aerogel structure rather than the large scale structure.

The coherence length varies with both temperature and pressure with the zero-temperature coherence length defined to be $\xi_0(P)=\left[ \frac{7\zeta(3)}{12}\right] ^{1/2}\frac{\hbar\medspace v_F(P)}{2\pi\medspace k_B T_c(P)}$, where $\zeta$ is the Riemann-zeta function, $v_F(P)$ is the pressure-dependent Fermi velocity, and $T_c(P)$ is the pressure dependent superfluid transition temperature. $\xi_0(P)$ varies from 15 to 80 nm between solidification pressure (34.4 bar) to 0 bar.  Most of the pressure dependence of $\xi_0(P)$ occurs between 0 and 6 bar. The experiments in Ref. \cite{zim.18} occur between 7.5 bar and 27 bar where $\xi_0(P)$ varies only from 34 to 18 nm. There are several different definitions for the temperature dependence of $\xi$. The most widely used definition is the Ginzburg-Landau (GL) correlation length given by:
\begin{equation}
\xi_{GL} (T) = \xi_0(P)(1-T/T_c)^{-1/2}.
\end{equation}
$\xi$ diverges near the second order phase transition and decays away with reducing temperature as $(1-T/T_c)^{-1/2}$. The GL coherence length is shown in Fig. \ref{fig:xi} for various pressures. 

When evaluated at $T_x$, $\xi$ at the various pressures all collapse into a narrow band of values around 50 nm, consistent with the model for the orbital-flop transition. The large variation of $T_x$ with pressure more or less converges to a narrow range of length scales. Above $T_x$, the $\xi$ is large so $\bm{\ell}$ is oriented perpendicular to the large scale structure. At the temperature when $\xi$ drops below roughly 50 nm, the orbital flop transition occurs and $\bm{\ell}$ is reoriented by the small scale structure. The crossover in scales seen in the correlation function and structure factor are of order $\sim$ 20-50 $r_0$. For silica aerogel with $r_0\sim$ 1.5 nm, this corresponds to order 30-75 nm compared with $\xi_{GL}(T_x,P) \sim 50$ nm. A more definitive test of this model would require going to lower pressure. At low pressure, the coherence length is substantially larger and we expect $T_x$ to drop in temperature as pressure is lowered. Below 1.5 bar, the zero-temperature coherence length is greater than 50 nm meaning no crossover transition is expected. 

Other experiments using a different type of planar aerogel also observe a phase diagram dominated by the A-phase with the orbital angular momentum orienting perpendicular to the planar sheets \cite{dmi.20}. However, an orbital flop transition was not observed in those experiments because the aerogel has much stronger anisotropy and does not appear to have the two different length scales. 

The sharpness of the orbital flop transition creates a useful experimental tuning parameter for probing new physics. Recently, it was shown that there is a substantial anomalous thermal hall effect in superfluid $^3$He in the presence of impurities like aerogel \cite{nga.20}. The direction of transverse thermal current is strongly dependent upon the orientation of the orbital angular momentum. Therefore, the orbital-flop can be used as a switch to turn on or off the transverse current. Because of the sharpness of the transition, the hall current should drop to zero abruptly as temperature is changed across $T_x$. This switching will be a definitive signature of the anomalous thermal hall effect. 

\section{conclusion}
In summary, we outline a procedure to simulate and characterize anisotropic aerogels with planar and nematic strands. The anisotropy is induced by biasing the diffusion process and can be characterized by the autocorrelation function, structure factor, and distribution of free paths. We make a connection to experimental aerogel by comparing the shape of the SAXS pattern with the structure factor. Both the calculated structure factor and the SAXS data exhibit a congruent dipolar shape at small-$\bm{q}$ and a perpendicular ellipsoidal pattern at large-$\bm{q}$. These two patterns reveal two different length scales of anisotropy in the aerogels. From this connection, we are able to classify real aerogel and show that stretched silica aerogel has large scale planar structure while compressed aerogel has large scale nematic structure. Finally, we provide a description of the aerogel's effect on the orbital angular momentum of superfluid $^3$He. The orbital angular momentum is oriented by the large scale structure in the aerogel at high temperature before spontaneously reorienting at a lower temperature due to the small scale structure. This "orbital-flop" transition can be leveraged in future work to observe the anomalous thermal hall effect in superfluid $^3$He.

This work was supported by the National Science Foundation, grant DMR-2210112.

\bibliography{Manref}

\clearpage
\section{Appendix}
\begin{figure}[h]
\includegraphics[width=1\textwidth]{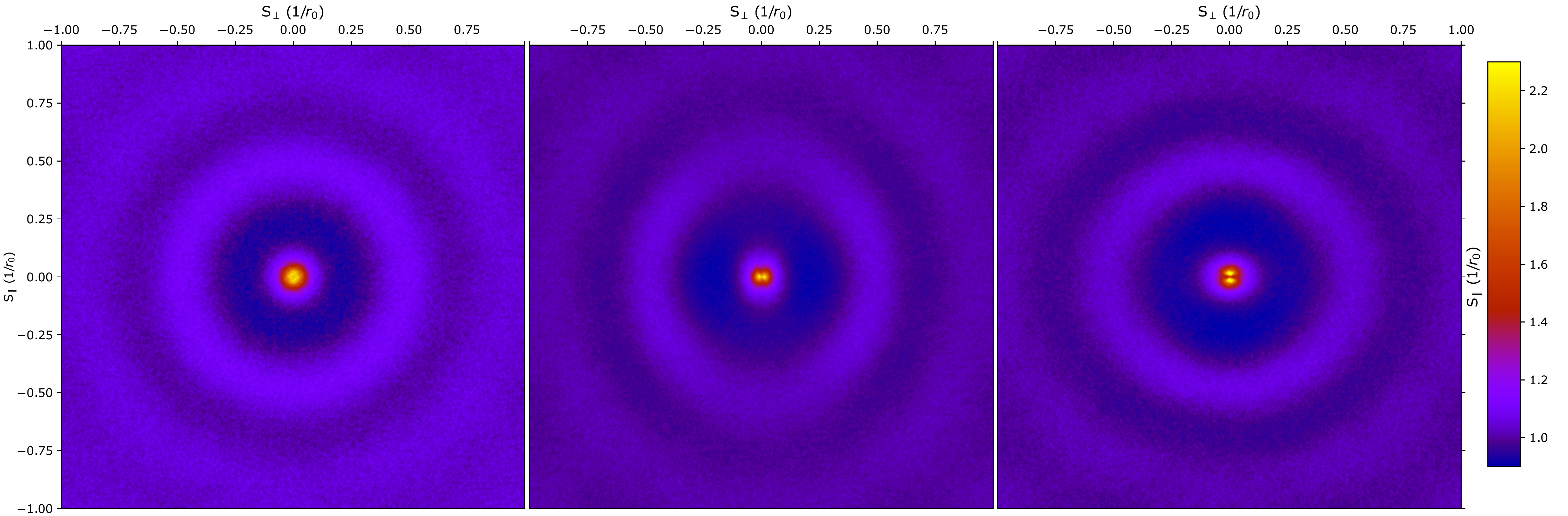}
\caption{The full structure factor of isotropic (left), nematic $\epsilon=0.125$ (center), and planar $\epsilon = 8$ (right) aerogels out to large $\bm{q}$.}
\label{fig:sq_full}
\end{figure}

The structure factors in Fig. \ref{fig:struct} show $S(\bm{q})$ out to $q \sim 0.1 \ r_0^{-1}$. While the small-angle X-ray scattering data is only dependent upon the small $\bm{q}$ behavior of $S(\bm{q})$, we can calculate the full structure factor out to $\bm{q}= \ r_0^{-1}$. As seen in Fig. \ref{fig:sq_full}, there are oscillations in the intensity at large $\bm{q}$ arising from the interparticle spacing. The differences in the two anisotropies are still evident at the smallest scale (largest $\bm{q}$). 

\begin{figure}
\includegraphics[width=1\textwidth]{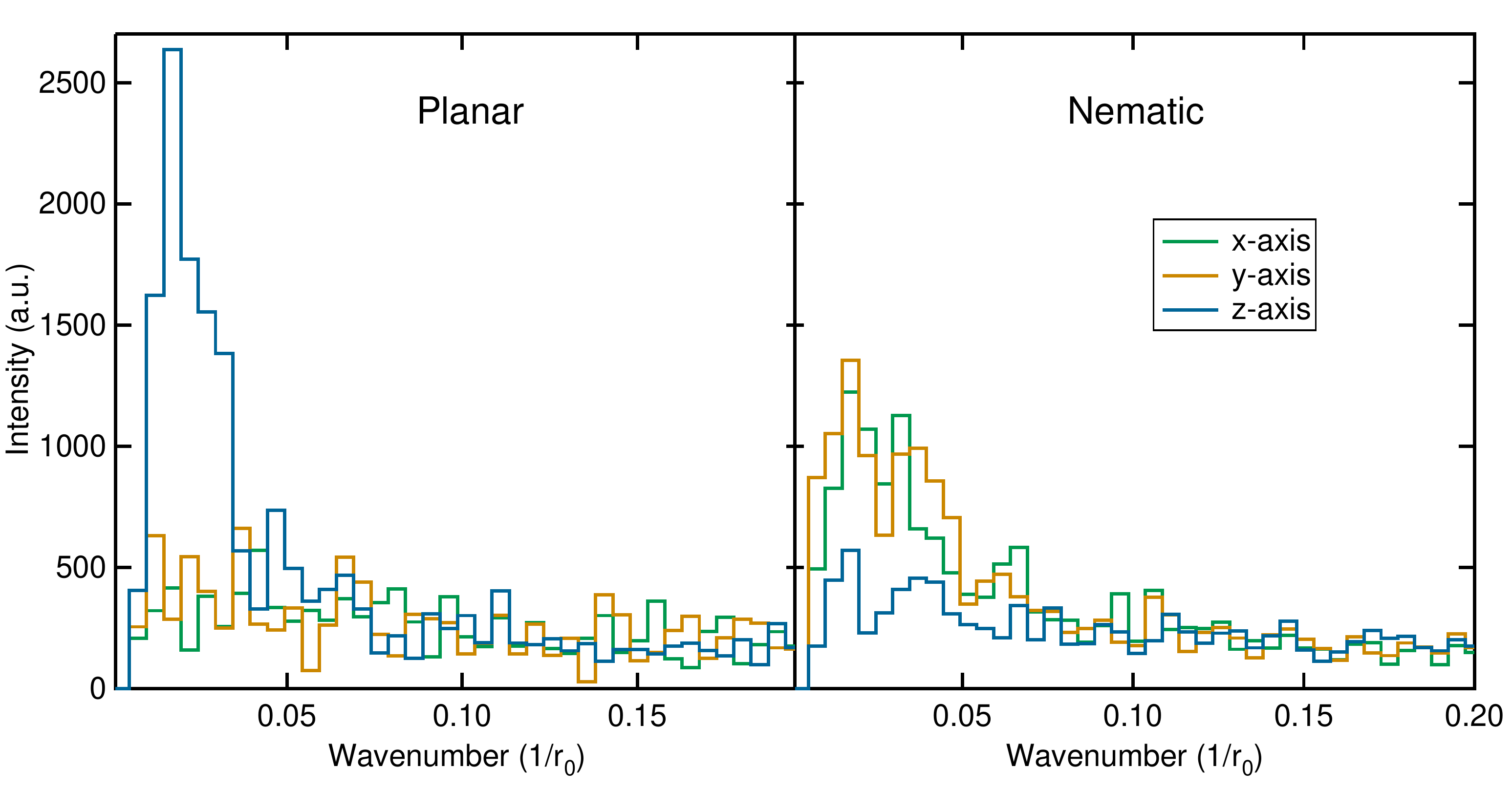}
\caption{Power spectrum of the density for planar $\epsilon=8$ (left) and nematic $\epsilon = 0.125$ (right) aerogels. For planar aerogels, there is a sharp peak around wavenumber $0.015 \ r_0^{-1}$. This gives a typical spacing between local maxima in density of about $60-70 \ r_0$. The size of the gaps between planes is then roughly half of that at $30 \ r_0$. There is not much density variation in the xy-plane. For nematic aerogel (right), there are peaks in the x- and y-axis density power spectrum which can be interpreted as the diameter of the nematic bundles.}
\label{fig:density_psd}
\end{figure}

\end{document}